\documentclass{aa}

\usepackage{graphicx}
\usepackage[varg]{txfonts}
\usepackage{hyperref}
\hypersetup{
    bookmarks=true,                                  % show bookmarks bar in pdf reader
    pdftitle={Revised age for CM Draconis},          % set pdf title          
    pdfauthor={Gregory A. Feiden, Brian Chaboyer},   % set pdf author
    colorlinks=true,                                 % false = box link, true = colored links
    linkcolor=blue,                                  % color of internal links
    citecolor=blue,                                  % color of citations
    urlcolor=blue                                    % external url color
}

\newcommand{\msun}{M_{\odot}}                        % Solar mass
\newcommand{\rsun}{R_{\odot}}                        % Solar radius
\newcommand{\teff}{T_{\rm eff}}                      % Effective temperature
\newcommand{\logg}{\log (g)}                         % log(surface gravity)
\newcommand{\wdcd}{WD~1633+572}                      % WD name
\newcommand{\water}{H$_2$O}                          % Water 

\begin{document}

   \title{Revised age for CM Draconis and WD 1633+572}

   \subtitle{Toward a resolution of model--observation radius discrepancies}

   \author{Gregory~A. Feiden\inst{1}
           \and
           Brian Chaboyer\inst{2}
          }

   \institute{Department of Physics and Astronomy, Uppsala University,
              Box 516, SE-751 20 Uppsala, Sweden. \\
              \email{gregory.a.feiden@gmail.com}
              \and
              Department of Physics and Astronomy, Dartmouth College,
              6127 Wilder Laboratory, Hanover, NH 03755, USA.
             }

  \date{Received 27 May 2014 / Accepted 16 September 2014}
 
  \abstract
    {We report an age revision for the low-mass detached eclipsing binary 
     CM Draconis and its common proper motion companion, \wdcd. An age of
     $8.5\pm3.5$~Gyr is found by combining an age estimate for the lifetime of
     \wdcd\ and an estimate from galactic space motions. The revised age is
     greater than a factor of two older than previous estimates. Our results 
     provide consistency between the white dwarf age and the system's galactic 
     kinematics, which reveal the system is a highly probable member of the 
     galactic thick disk. We find the probability that CM Draconis and \wdcd\ 
     are members of the thick disk is 8500 times greater than the probability 
     that they are members of the thin disk and 170 times greater than the 
     probability they are halo interlopers. If CM Draconis is a member of the 
     thick disk, it is likely enriched in $\alpha$--elements compared to iron
     by at least 0.2~dex relative to the Sun. This leads to the possibility 
     that previous studies under-estimate the [Fe/H] value, suggesting the system
     has a near-solar [Fe/H]. Implications for the long-standing discrepancies 
     between the radii of CM Draconis and predictions from stellar evolution
     theory are discussed. We conclude that CM Draconis is only inflated
     by about 2\% compared to stellar evolution predictions.}


   \maketitle
   
%
%________________________________________________________________

\section{Introduction}
\object{CM Draconis} (also GJ~630.1~AC; hereafter CM Dra) is of fundamental 
importance for understanding low-mass stellar structure and evolution. CM Dra 
is a detached double-lined eclipsing binary (DEB) consisting of two mid-M-dwarf 
stars \citep{Lacy1977} whose masses and radii are known with high precision. 
The primary has a mass $M_A = 0.23102 \pm 0.00089 \msun$ with a radius 
$R_A = 0.2534 \pm 0.0019 \rsun$ and the secondary has a mass 
$M_B = 0.21409 \pm 0.00083 \msun$ with $R_B = 0.2398 \pm 0.0018 \rsun$ 
\citep{Metcalfe1996,Morales2009a,Torres2010}. Both stars in CM Dra therefore 
have a mass below the nominal boundary where theory predicts stars to be 
fully convective throughout their interior \citep[$M \sim 0.35 \msun$;][]{
Limber1958a,CB97}.

The precision with which the fundamental properties of CM Dra are known
allows for direct verification of predictions from stellar evolution 
theory. One of most basic predictions of the theory is the stellar 
mass--radius relationship. Below the fully convective boundary, 
predictions from stellar evolution models are largely insensitive to physical
ingredients as the interiors are undergoing near--adiabatic 
convection. Variations of model predictions are typically at the 
1\% level, if provided with stringent constraints on the stellar 
metallicity and age. This caveat has typically been the limiting 
factor in rigorous tests of stellar evolution models, particularly 
for low-mass M-dwarfs \citep[e.g.,][]{Young2005,FC12,Torres2013}. 
CM Dra is one of only two fully convective systems for which strong 
constraints can be placed on the these properties. Recent metallicity
estimates converge toward a value of [Fe/H] $= -0.3\pm0.1$~dex
\citep{RojasAyala2012,Terrien2012,Kuznetsov2012} and an age of $4\pm1$ Gyr has been 
estimated from a white dwarf (WD) common proper motion companion, \wdcd\ 
\citep{Morales2009a}. 

At the given age and metallicity of CM Dra, stellar models are unable 
to accurately reproduce the observed radii \citep{FC14,MM14}. 
Models predict radii that are 6.0\% and 6.5\% too small for the primary 
and secondary, respectively. The commonly cited explanation for the 
discrepancies is the presence of strong magnetic fields and/or magnetic 
activity (i.e., spots). These mechanisms inhibit convective energy transport causing 
the stars to inflate as they attempt to maintain a constant energy flux through the 
surface \citep{MM01,Chabrier2007,MM11}.\footnote{We note that while spots are the physical
manifestation of the suppression of convection locally on the stellar photosphere, 
they are often considered separately from magneto-convection \citep{MM01,Chabrier2007} 
in 1D stellar evolution models.} 
Spots may also bias radius determinations from 
light curve data toward larger radii if present in an appropriate configuration,
namely spots clustered near the poles \citep{Windmiller2010,Morales2010}. However, we recently 
argued that, while magnetic fields may be 
able to cause such effects, the required internal magnetic field strengths are 
on the order of 1 -- 50 MG and are likely too strong to be stably supported 
within these stars \citep{FC14}, a conclusion also reached by \citet{Chabrier2007}.
Additionally, we argued that star spot properties needed to reconcile models and 
observations are not yet supported by empirical evidence and that
other avenues to reconcile models should therefore be explored.

Model assessments are predicated on the age and metallicity of CM Dra being 
correct and thus model discrepancies representing real departures of theory 
from observations. It has long been postulated that CM Dra is an old, Population
II object \citep[e.g.,][]{Lacy1977,CB1995}, but the WD age contradicts that 
assumption suggesting, instead, that the star is a Population I object. 
Here, we report a revision to the age of CM Dra based
on modeling its common proper motion companion, \object{WD 1633+572}, that is
supported by its galactic space motions. Section~\ref{sec:age} contains 
the derivation of the revised age, followed in Sect.~\ref{sec:metallicity} 
by arguments that the age and galactic kinematics suggest the system may have  
a near-solar metallicity. We then synthesize the metallicity and age results
in Sect.~\ref{sec:tage} and assess the impact of our results on the noted
radius inflation in Sect.~\ref{sec:model}. Finally, we provide a brief 
discussion of additional implications for the study in Sect.~\ref{sec:disc}.

\section{Age estimate}
\label{sec:age}

\subsection{White dwarf age}
\label{sec:wdage}
\wdcd\ is a DQ WD showing shifted C$_2$ Swan bands as well as C$_2$H absorption
\citep{Giammichele2012}. These peculiarities are thought to be characteristic of a 
He dominated atmosphere which has recently dredged up carbon from the bottom of the 
thin convective envelope \citep{Hansen2004}.
An age estimate for \wdcd, and therefore CM Dra, was previously provided by 
\citet{Morales2009a} who found the system to be $4.1\pm0.8$~Gyr old. Their
result relied on combing an estimate of the WD cooling age with an approximate
lifetime of the WD progenitor star. Cooling tracks predicted a WD cooling 
age of $2.84\pm0.37$~Gyr for a $0.63\pm0.04\msun$ WD \citep{Bergeron2001}.
Stellar evolution models provided a progenitor star lifetime of about $1.3$~Gyr 
assuming the progenitor star had a mass of $2.1\pm0.4\msun$ \citep{Catalan2008}. 

However, recent advances in WD atmosphere and cooling models lead to a downward revision 
of the mass for \wdcd. The new estimate is $M_{\rm wd} = 0.57\pm0.04\msun$ \citep{
Giammichele2012}. Decreasing the WD mass will increase the radius and 
decrease the WD cooling age, as a result. \citet{Giammichele2012} predict
the cooling age of \wdcd\ is 2.62~Gyr, slightly younger than before. An 
independent analysis of \wdcd\ confirms the mass and age estimate \citep[M. Salaris, priv.\
comm.;][]{Salaris2010}. Cooling tracks from \citet{Salaris2010} yield a WD
age of $3.4\pm0.6$~Gyr, consistent with previous estimates, within the error bars. 
This latter analysis allowed for phase separation during crystallization, which leads 
to an increase in the cooling age, and can account for about half of the difference
between the \citet{Salaris2010} and \citet{Giammichele2012} estimates.

Although the WD cooling age is not dramatically affected by updated cooling 
tracks, the lower mass estimate suggests that revision of the progenitor
mass---and age---is needed. There are multiple initial--final mass relations 
(IFMRs), each which predict different progenitor masses: $M_p = 1.5\pm0.5\msun$ 
\citep{Catalan2008}, $M_p = 1.3\pm0.4\msun$ \citep{Kalirai2009}, and 
$M_p = 1.6\pm0.9\msun$ \citep{Zhao2012}. Variation may be 
expected given that these studies sample different parameter regimes for WDs using 
different methods. The aforementioned relations all include WDs with masses around 
$0.57\msun$, but it should be noted that none contain WDs with a similar mass and 
$\teff$ as \wdcd. Instead of selecting a single IFMR upon which to base the progenitor
mass estimate, we take an uncertainty weighted average of those values listed
above and find $M_p = 1.4\pm0.3\msun$. This revised mass represents a significant 
reduction in the progenitor mass compared to \citet{Morales2009a} and translates 
into a significant age increase of the progenitor star lifetime.

Using standard Dartmouth stellar evolution models \citep{Dotter2008,FC14}, we 
evolve models at $1.1\msun$, $1.4\msun$, and $1.7\msun$ with a metallicity of 
$-0.3$~dex. Progenitor ages are taken to be the age of the 
model at the tip of the red giant branch (tRGB), as subsequent phases of 
evolution do not contribute appreciably to the overall progenitor age. From
these three models, we estimate the age of the progenitor star to be 
$3.1^{+3.8}_{-1.3}$~Gyr, where the errors represent the errors on the mean value.
The combined WD cooling age plus progenitor age for \wdcd\ is then $6.5^{4.4}_{-1.9}$
Gyr. However, this age is subject to further revision due to possible abundance 
revisions for CM Dra presented in Section~\ref{sec:metallicity}

\begin{figure*}[t]
	\begin{center}
		\includegraphics[width = 0.8\linewidth]{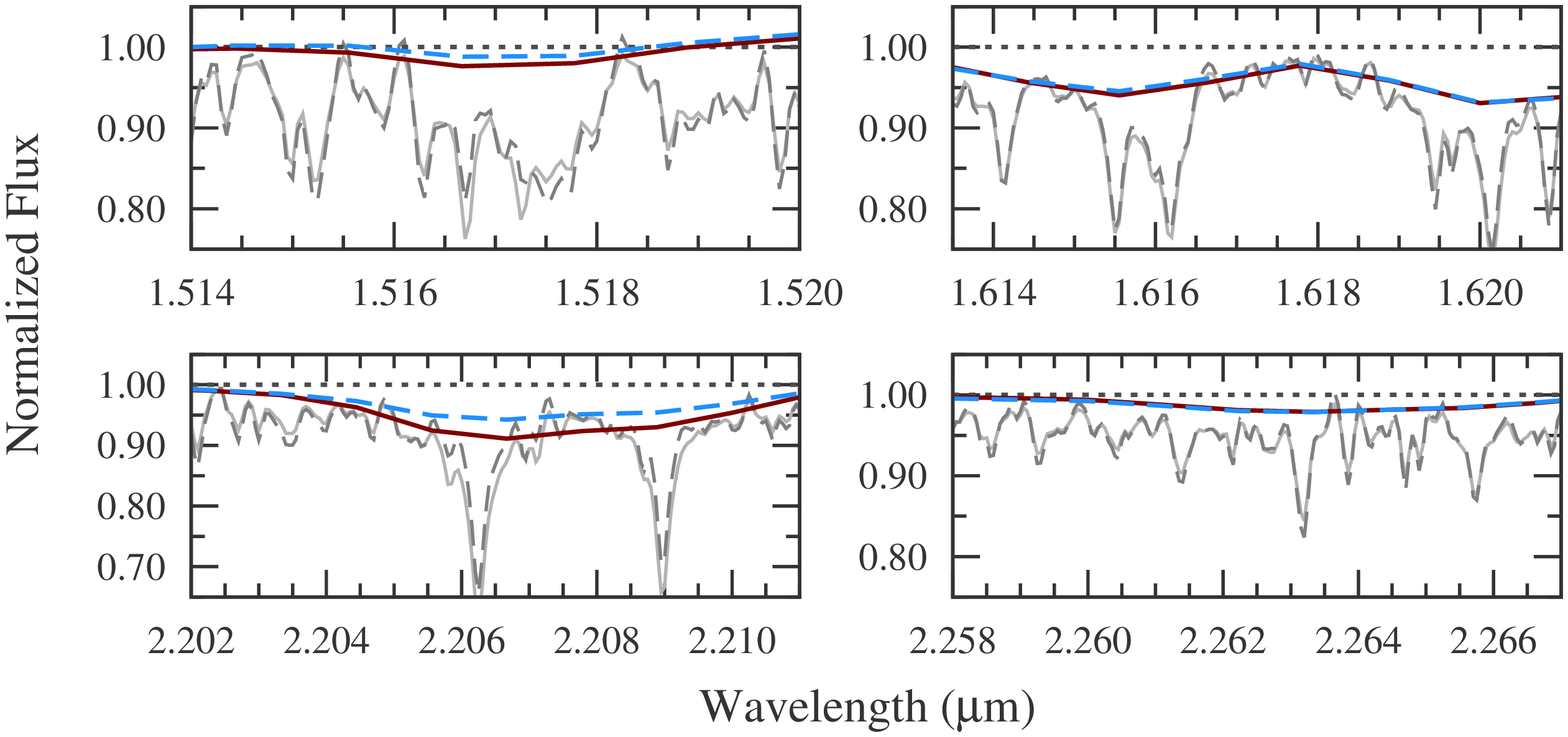}
	\end{center}
	\caption{Synthetic {\sc Phoenix BT-Settl} stellar spectra in the 
		four wavelength regions utilized in the \citet{Terrien2012a} NIR EW
		metallicity calibration. Spectra are shown for a star with $\teff = 3\, 200$~K,
		$\logg = 5.0$, and [Fe/H] = 0.0 with [$\alpha$/Fe] = 0.0 (maroon, solid 
		lines) and +0.2 (light-blue, dashed lines). These have been degraded to an
		approximate spectral resolution $R \sim 2\, 000$. For reference, the same
		spectra are shown with spectral resolution $R \sim 40\, 000$.
		\label{fig:specs}}
\end{figure*}

\subsection{Kinematic age}
Disagreement has been previously noted between the age of \wdcd\ and
the age inferred from the system's kinematics \citep{Morales2009a}.
The WD age is characteristic of the system belonging to the galactic 
thin disk population, while the system's high proper motion suggests a
kinematically older age with a possible Population II origin. 

CM Dra has proper motions of $\mu_\alpha = -1109$ mas yr$^{-1}$ and 
$\mu_\delta = 1203$ mas yr$^{-1}$ and \wdcd\ has commensurate 
values of $\mu_\alpha = -1106$ mas yr$^{-1}$ and 
$\mu_\delta = 1206$ mas yr$^{-1}$ \citep{LepineShara2005}. Parallaxes
of both objects were obtained in the US Naval Observatory parallax program
\citep{Harrington1980}. CM Dra has a parallax of $\pi = 68 \pm 4$ mas 
and \wdcd\ has a parallax of $\pi = 61 \pm 6$ mas, indicating the 
two systems are approximately equidistant from the Sun. The parallax of 
CM Dra has been revised to $\pi = 69.2 \pm 2.5$ mas by the Yale
Trigonometric Parallax program \citep{vanAltena1995}, but the Yale program 
did not provide a revision for \wdcd. Still, the parallaxes between
the two objects are consistent within the errors.

There are strong disagreements in the literature between the measured 
radial velocities (RVs) of
CM Dra and \wdcd. The absolute RV for CM Dra is a by-product of
accurate RV monitoring over 10 years in order to measure the masses of
the two stars \citep{Metcalfe1996,Morales2009a}. These authors find 
$\gamma = -118.24\pm0.07$ km $^{-1}$, although they admit that the uncertainty 
in this value is almost certainly larger. An independent study by \citet{Karatas2004} 
finds a value of $\gamma = -118.71$ km s$^{-1}$, confirming the RV measurement.
However, the absolute RV for \wdcd\ is quoted to be $\gamma = 3.4$ 
km s$^{-1}$ \citep{Silvestri2002,Sion2009,Sion2014}. Such a difference
in the RVs of CM Dra and \wdcd\ precludes the notion that they share
a common origin. 

Investigation of this issue revealed that the quoted RV for \wdcd\ 
was based on a single RV measurement of CM Dra \citep{Silvestri2002}. 
This was motivated by the fact that the two systems share a common 
high-proper motion and are, then, likely coeval. Therefore, the 
absolute RV of CM Dra could be safely projected onto \wdcd. However, it 
is likely that the authors did not realize that CM Dra was itself a tight 
M-dwarf binary. Relying on only a single spectrum of CM Dra likely 
produced an incorrect RV.  Knowing that the absolute RV of \wdcd\ was
assumed by the authors to be equal to that of CM Dra, it should be revised 
to $\gamma = -118$ km~s$^{-1}$, rescuing the assumption that the systems
share a common origin.

We find galactic space velocities of ($U$, $V$, $W$) = ($105 \pm 4$, $-120 \pm 1$, 
$-36 \pm 2$) km~s$^{-1}$ for CM Dra and ($U$, $V$, $W$) = ($119 \pm 9$, $-123 \pm 3$, 
$-31 \pm 4$) km~s$^{-1}$ for \wdcd, where the uncertainties are based only 
on the uncertainty in the measured parallaxes. Note that the sign of the $U$ 
velocity coordinate is with respect to galactic \emph{anti}-center. Corrections 
for the local standard of rest were not applied. 

While it is not possible to definitively associate any single star with a 
galactic population, it is possible to assign a relative probability that
a star belongs to a given stellar population. Assignment of relative probabilities
follows the procedure adopted by \citet{Bensby2005}. They assume galactic
space velocity distributions for the thin disk, thick disk, and stellar halo 
are Gaussian with resulting probabilities normalized to the observed fraction
of stars in the solar neighborhood \citep[see Appendix A of][]{Bensby2005,Bensby2014}. 
Given their velocity dispersions and asymmetric drifts for the three kinematic 
populations, we find that CM Dra is about 8500 times more likely to be a member
of the galactic thick disk than the galactic thin disk. Similarly, CM Dra is 170
times more likely to belong to the thick disk than the galactic halo. It
therefore appears statistically unlikely that CM Dra is a member of the thin 
disk and is better suited as a member of the thick disk. Although it is not
impossible to imagine a scenario whereby CM Dra is a highly perturbed thin disk
star, we shall provide further evidence that the properties of CM Dra are 
consistent with a possible thick disk membership.

We note that \citet{Sion2014} claim that all local WDs within 25~pc
are very likely members of the thin disk, including \wdcd. However, their 
membership claim is based on an erroneous RV measurement for CM Dra, which
leads to \emph{UVW} space velocities consistent with thin disk membership. With 
the correct RV value, it is very likely that \wdcd\ is a local member of 
the thick disk population.

Assignment to the thick disk population has consequences for both the age 
and chemical composition of CM Dra. It is thought that the thick
disk formed relatively early in the history of our Galaxy---about 11 Gyr 
ago---but star formation was truncated after $\sim$1 to 2 Gyr 
had elapsed. This scenario is supported by the imprint of 
various stellar population on chemical enrichment history of thick disk
stars \citep[e.g.,][]{Bensby2007,Bensby2014} and by age estimates some of the
oldest WDs in globular clusters and the solar neighborhood, which are 
between 9 and 12 Gyr \citep{Hansen2013, Salaris2010}, for an average
of about $10.5\pm1.5$~Gyr. 
There is also strong evidence suggesting that thick disk stars have a range
of metallicities, from metal-poor up to solar values \citep{Bensby2007}. 
However, thick disk stars can be distinguished from thin disk stars by the 
fact that they appear enriched in $\alpha$-elements \citep{Bensby2010,Bensby2014,
Adibekyan2013}. At [Fe/H] $= -0.3$~dex, thick disk stars are characterized 
by [$\alpha$/Fe] $\sim$ $+0.2$~dex to $+0.4$~dex. If CM Dra is a member of
the thick disk population, we may infer that it is $\alpha$-enhanced with 
an age of $10.5\pm1.5$~Gyr old.

\begin{table*}
	\caption{Metallicity determinations for {\sc Phoenix BT-Settl} spectra calculated using a NIR EW calibration.}
	\label{tab:metals}
	\centering
	\begin{tabular}{l l l l l l l l l}
		\hline\hline\noalign{\smallskip}
		$\teff$   &  \multicolumn{2}{c}{[$\alpha$/Fe] = +0.0} & & \multicolumn{2}{c}{[$\alpha$/Fe] = +0.2} \\
		          \noalign{\smallskip}\cline{2-3}\cline{5-6}\noalign{\medskip}
		(K)   &   [Fe/H]$_{H}$   & [Fe/H]$_{K}$ & &  [Fe/H]$_{H}$   & [Fe/H]$_{K}$ & $\Delta$[Fe/H]$_{H}$
		          & $\Delta$[Fe/H]$_{K}$ \\
		\noalign{\smallskip}\hline\noalign{\smallskip}
		3000 & $+0.19$ & $-0.23$ & & $-0.10$ & $-0.49$ & $-0.29$ & $-0.27$ \\
		3100 & $+0.25$ & $-0.24$ & & $+0.01$ & $-0.48$ & $-0.24$ & $-0.24$ \\
		3200 & $+0.31$ & $-0.27$ & & $+0.13$ & $-0.47$ & $-0.18$ & $-0.20$ \\
		3300 & $+0.38$ & $-0.27$ & & $+0.27$ & $-0.44$ & $-0.11$ & $-0.17$ \\
		\noalign{\smallskip}\hline
	\end{tabular}
\end{table*}

\section{Impact on metallicity determination}
\label{sec:metallicity}
Assuming that CM Dra is a member of the galactic thick disk, and thus $\alpha$-element
enriched, has implications for M-dwarf metallicity determinations based on calibrations of
near-infrared (NIR) equivalent widths (EWs) \citep[e.g.,][]{Terrien2012a,
RojasAyala2012}. These calibrations are performed on wide binaries in the solar 
neighborhood with an FG primary and an M-dwarf companion, where the metallicity
of the FG primary is measured and projected onto the M-dwarf. Since the binaries
are in the solar neighborhood, the calibration sample is biased toward thin disk 
stars with solar-like distributions of heavy elements. Increasing [$\alpha$/Fe], 
particularly [O/Fe], increases the level of continuum suppression caused by \water\ 
molecules in the NIR at a given [Fe/H]. As a consequence, NIR atomic line depths 
will appear weaker with respect to a normalized pseudo-continuum than in the case
of a solar-like metal distribution. Metallicity determinations based on EWs of 
NIR atomic lines are then expected to produce [Fe/H] values that are too low 
when applied to a star that is \emph{unknowingly} $\alpha$-enhanced compared 
to the Sun. 

To test the influence of an enhanced [$\alpha$/Fe] on the abundance determination
of CM Dra, we applied the metallicity calibration of \citet{Terrien2012a} to a 
series of {\sc Phoenix BT-Settl} synthetic spectra \citep{Allard2012}. The spectra had [Fe/H] = 0.0 
with a \citet{Caffau2011} solar composition, $\logg = 5.0$, and ranged in $\teff$ 
from $3\,000$~K to $3\,300$~K. There were two sets of spectra with these parameters,
one set with [$\alpha$/Fe] = 0.0 and the other set with [$\alpha$/Fe] = 0.2. 
Since the original EW calibration was performed using spectra with $R \sim 2\,000$,
we degraded the synthetic spectra by convolving them with a Gaussian kernel. 
Examples of the resulting spectra at $3\,200$~K in the wavelength regions used 
for the \citet{Terrien2012a} metallicity determination are shown in Fig.~\ref{fig:specs}. 
For reference, Fig.~\ref{fig:specs} also shows the same spectra degraded to 
$R \sim 40\,000$, where atomic features are more easily identified.

Applying the \citet{Terrien2012a} [Fe/H] calibration to each spectrum in the series,
we find that $\alpha$-enhanced spectra yield [Fe/H] values that are systematically 
lower by 0.1~dex -- 0.3~dex than spectra with a solar $\alpha$-abundance. Results of 
this analysis are tabulated in Table~\ref{tab:metals}. This difference is independent 
of the degraded spectral resolution and whether one uses the $H$-- or $K$--band relation, 
but it is dependent on $\teff$.
Differences increase with decreasing $\teff$. Temperature dependence is expected as 
\water\ absorption increases with decreasing $\teff$ at constant metallicity, a trend 
that is supported by empirical data \citep{RojasAyala2012}. Therefore, increasing the 
relative abundance of \water\ will have a larger impact at cooler $\teff$, as was found
in our analysis.

From Table~\ref{tab:metals}, we see that the metallicity calibration does not successfully 
reproduce the [Fe/H] value of the synthetic spectra. This is likely a reflection of both 
uncertainties in synthetic model atmospheres and slight differences in continuum 
normalization. We stress that this is not a reflection of the intrinsic quality of 
the metallicity calibration. \citet{RojasAyala2012} demonstrated that the $K$-band 
\ion{Na}{I} doublet is weaker in {\sc BT-Settl} spectra than in empirical data. Additionally,
they showed that {\sc BT-Settl} spectra possess weaker \ion{Ca}{I} triplet lines in 
the relevant $\teff$ range. This suggests one would derive lower [Fe/H] values from 
synthetic spectra. However, at shorter wavelengths, \ion{Ca}{I} features appear stronger 
in {\sc BT-Settl} spectra. Assuming that \ion{Ca}{I} lines continue to appear stronger in 
the $H$-band, we would expect to find higher [Fe/H] values returned from the $H$-band 
calibration. These trends are consistent with offsets in the [Fe/H] determinations listed 
in Table~\ref{tab:metals} at solar [$\alpha$/Fe]. Curiously, we note that the average abundance 
between the $H$-- and $K$--band is accurate.

Although the absolute [Fe/H] determination is suspect when applied to synthetic spectra, 
these errors will be mitigated in a relative abundance study to assess the impact of 
$\alpha$-enhancement. Additional errors may be present in the strength of \ion{Ca}{I} 
features since Ca is an $\alpha$-element, but there should be no relative impact on the
strength of Na or K atomic features. Looking at the $\Delta$[Fe/H] values in Table~\ref{tab:metals},
one sees the differences between metallicity errors introduced by $\alpha$-enhancement are 
consistent between $H$--band calibration results (equally dominated by the EW of \ion{Ca}{I} and 
\ion{K}{I}) and $K$--band calibration results (dominated by the Na doublet). Errors due to the
additional abundance of \ion{Ca}{I} do not appear to strongly affect the results. 
 
We conclude that CM Dra may have an [Fe/H] about 0.2~dex higher than quoted by 
\citet{Terrien2012}, assuming a $\teff \approx 3200$~K.
This implies [Fe/H] = $-0.1\pm0.1$~dex with [$\alpha$/Fe] = 0.2 dex. Since CM Dra and 
\wdcd\ are assumed to have a common origin, the progenitor of \wdcd\ should be modeled
with the same abundance. Increasing the overall metallicity of the progenitor will act
to increase its lifetime and therefore increase the age of the system.

\section{Final age}
\label{sec:tage}
The association of CM Dra and \wdcd\ with the galactic thick disk leads 
to a revision of the properties of the WD progenitor. Stellar models 
must be computed with [$\alpha$/Fe] $= +0.2$~dex and assuming [Fe/H] $= -0.1$~dex. 
Computing models at the same progenitor masses as in Section~\ref{sec:age}
yields a progenitor age estimate of $3.8^{+4.8}_{-1.4}$~Gyr. \wdcd\ then 
appears to have an age of $7.2^{+5.4}_{-2.0}$~Gyr. If we assume that the 
oldest age must correspond to the maximum age of the thick disk (about 12 Gyr),
then the age may be written $7.2^{+4.8}_{-2.0}$, or $8.5\pm3.5$~Gyr if all ages
in this range are equally likely \emph{a priori}. This age is consistent with
previous estimates, within the error bars, but represents nearly a factor of two 
increase in the mean value.

Furthermore, updates to WD atmosphere models, cooling models, and the IFMR lead 
to consistent age estimates 
between WD age-dating of \wdcd\ and the association of CM Dra and \wdcd\ with
the galactic thick disk population via kinematics. The WD age is still highly
uncertain (within 30\%) due to uncertainty in the IFMR and thus an uncertainty
in the progenitor star lifetime. However, considering the minimum age for a 
thick disk member is in the vicinity of 9~Gyr, the WD age provides good
agreement. There exists the question of whether a 9 -- 12 Gyr old thick disk
object can be characterized by [Fe/H] $= -0.1$ while still showing signatures 
of $\alpha$-enhancement. There is evidence that FGK stars with this abundance
pattern exist \citep{Bensby2007,Bensby2010,Bensby2014,Adibekyan2013}, where 
the more metal-rich thick disk members are the product of subsequent star formation
and enrichment episodes. \citet{Bensby2007} suggest that chemical enrichment
of thick disk stars carried on until about 8 or 9 Gyr ago, which is largely consistent 
with the age derived for CM Dra and the finding that it possesses a 
significantly high metallicity for a thick disk star. We infer that a probable 
age for CM Dra is anywhere between 8 and 12 Gyr, given its likely association 
with the thick disk. For modeling purposes, we will avoid constraining the age
further and adopt the formal age range of $8.5\pm3.5$~Gyr.

\section{Implications for stellar evolution theory}
\label{sec:model}
Dartmouth stellar evolution models are used to assess the impact on both 
standard stellar models \citep{Dotter2008,FC14} and those that include 
magnetic field effects \citep{FC12b}.

\subsection{Standard stellar evolution models}
\begin{figure}[t]
	\begin{center}
		\includegraphics[width = 0.9\linewidth]{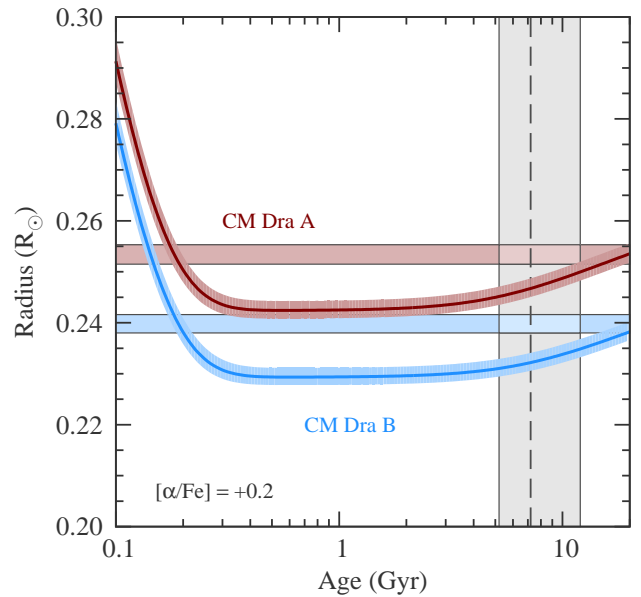}
	\end{center}
	\caption{Standard Dartmouth models computed at the precise masses of the CM Dra stars 
		with [Fe/H] $= -0.1\pm0.1$~dex with [$\alpha$/Fe] $= +0.2$~dex. Solid 
		lines show the evolution of models with [Fe/H] $= -0.1$~dex and the 
		band surrounding those tracks show the predicted variation with metallicity. 
		The lower bound of the uncertainty band corresponds to [Fe/H] $= -0.2$~dex, 
		while the upper band corresponds to solar metallicity. For reference, the 
		horizontal shaded regions mark the observed radius with $1\sigma$ uncertainties. 
		The vertical stripe denotes the age constraint determined in Section~\ref{sec:tage},
	 	with a vertical dashed line marking an age of 7.2~Gyr. \label{fig:cmdra_std}}
\end{figure}

Figure~\ref{fig:cmdra_std} provides a comparison between standard stellar
evolution model predictions and the observed properties of CM Dra. 
After about 5~Gyr, evolutionary effects become noticeable and model radii increase 
over their zero-age-main-sequence value. At the former age of 4~Gyr, the model radii have 
not evolved significantly, compared to the zero-age-main-sequence values, and 
are about 6\% discrepant with observations.
However, at 7.2~Gyr, models of the primary shown in Fig.~\ref{fig:cmdra_std} 
have radii of $0.2450\rsun$, $0.2467\rsun$, and $0.2487\rsun$ for [Fe/H] 
= $-0.2$, $-0.1$, and 0.0, respectively. These correspond to relative radius errors
of 3.3\%, 2.6\%, and 1.9\%, respectively, or deviations at the $4.4\sigma$, 
$3.5\sigma$, and $2.5\sigma$ level. Similarly, for the secondary, models
have radii at 7.2~Gyr of $0.2306\rsun$, $0.2322\rsun$, and $0.2341\rsun$ for the
same metallicities, respectively. Discrepancies with the 
secondary are larger at 3.8\%, 3.2\%, and 2.4\%, or $5.1\sigma$, $4.2\sigma$,
and $3.2\sigma$, respectively. Notably, while the radii are incorrectly predicted, 
the effective temperature ratio between the two models agrees with the observed value
within the error bars. 
For [Fe/H] = $-0.1$~dex, $\teff = 3297$~K and $3270$~K for the
primary and secondary, respectively, giving a temperature ratio equal to 0.9918, effectively
consistent with the observed value of $0.996\pm0.004$ \citep{Morales2009a}.
While increasing the age and overall metal abundance of the stellar models helps to relieve 
the size of the radius discrepancies, significant ($3\sigma$) discrepancies remain. 

If we instead assume that CM Dra is formally a member of the thick disk, and thus nominally 
between the age of 8 and 12 Gyr, the discrepancies between models and observations is further
lessened. From Fig.~\ref{fig:cmdra_std}, we find that a model of the primary has a radius 
$R = 0.2493\rsun$ for [Fe/H] = $-0.1$~dex at 10 Gyr yielding a relative radius error
of 1.6\% ($2.2\sigma$). A model of the secondary has a radius of $0.2339\rsun$,
which is 2.5\% ($3.3\sigma$) discrepant with the observations. The temperature ratio remains
equal to 0.9918, as above. One can further reduce the 
discrepancies by assuming an older age. Nevertheless, the models can not be completely reconciled
with the observations for ages less than the age of the Universe.

\subsection{Magnetic stellar evolution models}

\begin{figure}
	\begin{center}
		\includegraphics[width = 0.9\linewidth]{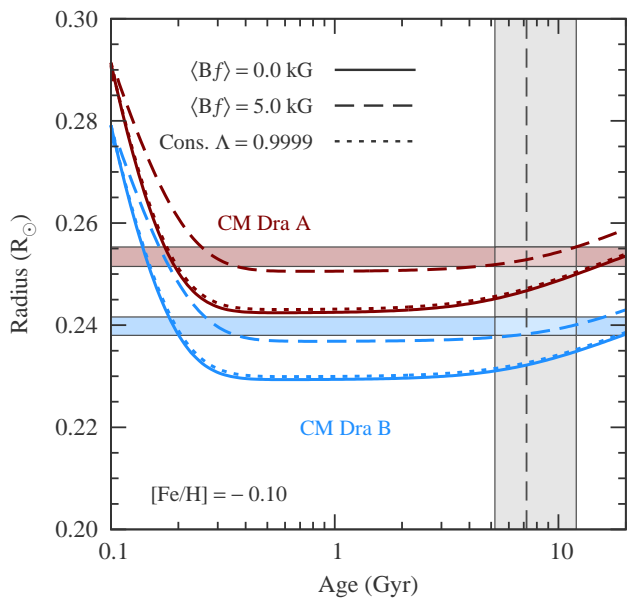}
	\end{center}
	\caption{Magnetic stellar evolution model tracks of the CM Dra stars.
	    Models are computed at the precise masses of the CM Dra stars 
		with [Fe/H] $= -0.1$~dex and [$\alpha$/Fe] $= +0.2$~dex. Solid 
		lines show the evolution of models with no magnetic field, for reference.
		Magnetic models with a ``rotational dynamo'' and dipole radial profile
		are shown as dashed lines while dotted lines are magnetic models computed
		with a constant-$\Lambda$ radial profile with a ``turbulent dynamo.''
		Horizontal shaded regions mark the observed radius with $1\sigma$ uncertainties. 
		The vertical stripe denotes the age constraint determined in Section~\ref{sec:tage}.
	 	\label{fig:cmdra_mag}}
\end{figure}

Stellar model mass tracks with magnetic fields are shown in Fig.~\ref{fig:cmdra_mag}.
Two formulations of the influence of magnetic fields are shown: a rotational dynamo with 
a dipole radial magnetic field strength profile, and a constant--$\Lambda$ turbulent dynamo 
\citep[see][for details]{FC14}. In general, the turbulent dynamo formulation does not 
provide an adequate solution, whereas a rotational dynamo with a 5.0~kG surface magnetic 
field strength is able to provide agreement. In addition to matching the observed stellar
radii, the rotational dynamo models are able to maintain agreement with the observed 
effective temperature ratio between the two components. 
Primary and secondary $\teff$s are reduced to $3032$~K and $3015$~K, respectively, 
for a ratio of 0.9944.
As in \citet{FC14}, the interior magnetic
field strength peaks at about 50 MG, making the magnetic field buoyantly unstable. 
However, by shifting the radial location of the peak interior magnetic field strength 
from $R = 0.15\ R_{\star}$ to $R = 0.50\ R_{\star}$, we are able to reduce the peak
magnetic field strength from 50 MG to 50 kG, for the quoted 5.0~kG surface magnetic
field strength, while still maintaining agreement between the models and observations 
between 7.2 and 12.0 Gyr. This revised value is more inline with expected magnetic 
fields strengths from 3D magnetohydrodynamic models \citep{Chabrier2006,Browning2008}.

\citet{FC14} discussed that the surface magnetic field has little effect on the radius 
inflation in fully convective stars, particularly CM Dra. However, the reduction in the 
model-observation radius discrepancies from 6\% to 3\% means that the model does not need
to develop a radiative core or shell through stabilization of convection. Surface
effects appear more able to correct the discrepancies when they are reduced to a few 
percent. We note that a 5.0~kG average surface magnetic field is stronger, by about a factor 
of two, than the typical value of 3~kG observed on fully convective stars, again raising
questions as to the validity of the magnetic models for these stars 
\citep[e.g.,][]{Reiners2012a}.

To reduce the surface magnetic field strength, we must instead compensate with larger 
interior magnetic fields. Thus, although the desired radius inflation is relatively small 
compared to previous studies, models still require a magnetic field strength of around 1~MG, 
which is at risk of being buoyantly unstable.
This requirement is unavoidable in current 1D magneto-convection prescriptions that 
aim to stabilize the stellar interior against convection. A simple order of magnitude
estimate illustrates this fact. Convection in fully convective stars is nearly-adiabatic,
with $\nabla - \nabla_{\rm ad} \sim 10^{-6}$. Magneto-convection prescriptions used by 
\citet{MM11} and \citet{FC12b} depend on $\nabla - \nabla_{\rm ad}$ being larger
than some value $\epsilon$, which is primarily set by the ratio of the magnetic pressure 
to the gas pressure. Therefore, convection will be suppressed when $\epsilon \sim 10^{-6}$.
Gas pressure deep in the stellar interior is $\sim 10^{16}$ to $10^{17}$ dyne~cm$^{-2}$,
meaning the magnetic pressure must be $\sim 10^{10}$ to $10^{11}$ dyne~cm$^{-2}$. This 
corresponds to a magnetic field strength of about $10^6$~G. One way out of this
requirement is through a turbulent dynamo, but as Fig.~\ref{fig:cmdra_mag} shows, 
more advanced techniques need to be explored if it is to impart significant structural
changes.

\subsection{Star spots}
We can, instead, consider star spots to be the source of the observed inflation 
\citep{Chabrier2007,Morales2010}, due to both the biasing of radius measurements
and actual structural changes inflicted on the stars. Taking both effects into 
account, 10\% of the surface would need to be covered in completely dark spots.
Assuming only a radius bias or only structural changes lead to 16\% and 25\%
coverages, respectively. Since spots are not necessarily completely dark, but 
simply cooler than the surrounding photosphere, adopting a temperature contrast 
of 90\% \citep{Berdyugina2005,FC14} leads to equivalent surface coverages of 30\%, 
46\%, and 64\%, for the aforementioned cases, respectively. For radius biasing to 
occur, which has a more substantial influence than structural changes, spots are 
required to be located preferentially at the poles. In principle, spots can be 
detected by detailed modeling of spectral line profiles, with polar cap spots often 
providing clear signal of their presence \citep[e.g.,][]{Berdyugina2005}. An 
observational search for polar cap spots on CM Dra is underway.

Assuming that a fraction of the stellar surface is covered in dark spots, one can
also estimate the expected level of light curve modulation. Numerical experiments
were carried out by \citet{Morales2010} to investigate this relationship. With 10\%
surface coverage, they found light curve modulation amplitude in the $R$-band was 
typically greater than 4\%, with the exact number depending on the distribution of
spots and the assumed spot sizes. CM Dra was observed to have a modulation amplitude
in the $R$-band of 3\% \citep{Morales2009a}. It is therefore conceivable that the 
light curve modulation could be fit assuming a 10\% coverage fraction. A more complete 
spot analysis exploring the entire parameter space would need to be performed, but 
is beyond the scope of this study. However, the key fact is that based on the work
of \citet{Morales2010}, the spots must be preferentially located at the poles.

\section{Discussion}
\label{sec:disc}
\subsection{Orbital eccentricity}
A rather curious feature of the CM Dra system is its mild orbital eccentricity 
\citep[$e = 0.05$;][]{Metcalfe1996,Morales2009a}. In its present configuration, 
the stars in CM Dra are expected to synchronize their rotation and circularize 
their orbit within roughly 0.3~Gyr \citep{Zahn1977}. Thus, the presence of an
elliptical orbit is rather perplexing. Though \wdcd\ could cause perturbations
to the orbit of CM Dra, at a projected distance of roughly 370~AU, the overall
impact on the orbit over time is likely to be negligible owing to the short orbital
period of CM Dra \citep{Morales2009a}. It has therefore been proposed that
CM Dra is host to a fourth body in the form of either a massive planet or
low-mass brown dwarf \citep{Deeg2008,Morales2009a}, though evidence for such a
companion has yet to be firmly established.

Another suggested interpretation might be that CM Dra is in fact younger than it 
appears, not having time to circularize its orbit. However, we know that \wdcd\
is at least 2.8~Gyr old, based on the cooling tracks alone and ignoring the progenitor
lifetime. Circularization, it was mentioned, should occur for this system within
roughly 0.3 Gyr. Though the tidal circularization calculation is only an order of 
magnitude estimate, it suggests that a younger age would not provide a sufficient
explanation for the observed eccentricity, unless of course, \wdcd\ and CM Dra do
not have a common origin and are instead the product of a stellar encounter.

\subsection{Mass--radius problem for low-mass stars}
Revisions to the properties of CM Dra advanced in this paper are unable to provide
a complete solution to the mass--radius discrepancies. However, the disagreements
are significantly reduced from of order 6\% to between 2\% and 3\%, further highlighting the 
need for accurate metallicities and reasonable age constraints if comparisons with
stellar evolution models are to be meaningful \citep{Young2005,FC12,Torres2013}. 
Additional uncertainties introduced by unknown He abundances complicate the matter
\citep{Valcarce2013}. Increasing the He abundance of the stars in CM Dra to a value
of $Y = 0.35$ (compared to $Y = 0.28$ in aforementioned models) can produce a 2\% 
radius increase, and thus lead to agreement between observations and theory. 
Subpopulations of stars in a few globular clusters show evidence of having 
significantly enhanced helium abundances ($Y \sim 0.35$ -- $0.40$), as in \object{NGC 2808}
\citep{Milone2012,Marino2014} and \object{Omega Centauri} \citep{Bellini2010,Dupree2011,Dupree2013}.
Presently, there is no evidence for such helium enhancement among field stars in
the halo or thick disk and models that can explain the multiple populations observed
in globular clusters are not applicable to field stars. Thus, there is no observational
evidence or theoretical arguments to support the idea that CM Dra may be significantly
enhanced in helium. Furthermore, such a prediction is at risk of being observationally
untestable.

Decreasing disagreements to the 2\% level does raise the question, ``How accurate
can one expect stellar evolution models to be?'' The level of uncertainty in stellar
models is often around 1\%, given current microphysics, making 2\% deviations 
potentially significant. The significance of the deviations are reinforced by the
fact that no well characterized fully convective stars appear smaller than models,
indicating a systematic offset. On the other hand, this also highlights the need for 
observers to mitigate systematics, especially those from spots that may be 
introducing errors on the order of 1 to 3\% \citep{Windmiller2010,Morales2010}.
Studies aimed at increasing the number of fully convective stars with empirically
determined star spot maps would provide a valuable contribution.

\subsection{Confirming $\alpha$-enrichment}
The hypothesis that CM Dra is a thick disk member requires confirmation. Evidence
presented throughout this paper, we believe, lends strong support to the idea, but 
the evidence presented is circumstantial. Identifying spectral signatures of 
$\alpha$-enhancement would provide the strongest evidence, as $\alpha$-enhancement
leads to the notion that the metallicity is near-solar. Investigations are ongoing
to identify unambiguous signatures in optical and NIR spectra. 

One further implication of CM Dra being $\alpha$-enhanced is that it may possibly 
reconcile disagreements among various metallicity estimates, which are well-documented 
\citep[see, e.g.,][and references therein]{Terrien2012}. Molecular features in optical
spectra leads to near-solar metallicities \citep{Gizis1997}, as do NIR photometric 
colors \citep{Leggett1998}. However, line modeling of atomic features and CO bands in 
the NIR yield consistently lower metallicities around $-0.6$ to $-1.0$~dex \citep{Viti1997,
Viti2002,Kuznetsov2012}, whereas NIR EW calibrations \citep{RojasAyala2012,Terrien2012} 
and NIR photometric relations \citep{Johnson2009} yield more intermediate values, as discussed 
earlier. Synthetic spectra that are $\alpha$-enhanced display molecular features in 
the optical that are similar to non-$\alpha$-enhanced spectra, whereas in the NIR, 
additional continuum suppression can lead to weaker atomic features, as well as weaker
CO features. The former would lead one to a more correct metallicity, while the
latter occurrences would provide lower metallicities. It has been previously suggested
that CM Dra is chemically peculiar \citep{Viti1997,Viti2002}, so perhaps this is the
manifestation of $\alpha$-enhancement. We are continuing to investigate this 
possibility.

\begin{acknowledgements}
The authors thank the anonymous referee for giving helpful comments and suggestions
to improve the manuscript and M. Salaris for providing a cooling track analysis of the 
white dwarf.
G.A.F.\ thanks B.\ Gustafsson, O.\ Kochukhov, E.\ Stempels, and T.\ Nordlander for 
stimulating discussion and helpful suggestions. The Dartmouth magnetic stellar evolution
code was developed with support from the National Science Foundation (NSF) grant 
AST-0908345. This work made use of NASA's Astrophysics Data System (ADS) and the 
SIMBAD database, operated at CDS, Strasbourg, France.
\end{acknowledgements}

% Bibliography using bibtex & natbib w/ ApJ style formatting
\bibliographystyle{apj}

\end{document}